\pdfoutput=1

\documentclass{article}

\usepackage{hyperref}

\usepackage{amssymb}
\usepackage{graphicx,amssymb,amsmath}
\usepackage[utf8]{inputenc}
\usepackage{authblk}
\usepackage[usenames, dvipsnames]{color}
\usepackage[utf8]{inputenc}
\usepackage[english]{babel}
\usepackage{amssymb}
\usepackage{comment}
\usepackage{amsthm}
\usepackage{caption}
\usepackage{algorithmic}
\usepackage{subfig}
\usepackage{enumitem}
\usepackage[export]{adjustbox}

\usepackage[algoruled]{algorithm2e}
\SetKwInOut{Input}{input}
\SetKwInOut{Output}{output}
\SetKw{KwT}{True} 
\SetKw{KwF}{False}
\SetKw{KwOr}{or}
\SetKw{KwAnd}{and}
\DontPrintSemicolon
\newtheorem{theorem}{Theorem}
\newtheorem{lemma}{Lemma}

\newtheorem{obs}[lemma]{Observation}

\graphicspath{{figures/}}
 
\newcommand{\etal} {{\it et al.}\xspace}

\definecolor {name} {rgb} {0.5,0.0,0.0}

\begin{document}


\title{Maximum-Area quadrilateral in a Convex Polygon, Revisited} 

\author[1]{Vahideh Keikha}
\affil[1]{\small Department of Mathematics and Computer Science, Amirkabir University of  Technology, Tehran, Iran}


\author[2]{Maarten L\"offler}
\affil[2]{\small Department of Information and Computing Sciences, Utrecht University, Utrecht, The Netherlands}
\author[1]{Ali Mohades}

\author[2]{J\'er\^ome Urhausen}
\author[2]{Ivor van der Hoog}

\maketitle

\begin{abstract}
In this note we show by example that the algorithm  presented in 1979 by Dobkin
and Snyder~\cite{45} for finding the largest-area $k$-gon that is inscribed in a convex polygon fails to find the optimal solution for $k=4$.The question whether the algorithm works when $k=4$ was posed by Keikha et al in~\cite{kluv} where they showed that the Dobkin Snyder algorithm fails for $k=3$. In this note we show that this problem and the problem for any constant value of $k$  can be solved in linear time.

\end{abstract}
This paper has been merged with \url{https://arxiv.org/abs/1705.11035}.











\section{Introduction}

Surprisingly, in~\cite{kluv} the authors show that the linear-time algorithm presented in 1979 by Dobkin
and Snyder~\cite{45} for finding the largest-area triangle that is inscribed in a convex polygon fails to find the optimal solution.   Also, Boyce \etal~\cite {48} and Avis \etal~\cite{mmud} observe that the algorithm by Dobkin and Snyder fails for $k\geq 5$ and $k=2$, respectively. What remains is to show that the algorithm fails for $k=4$.

\begin {figure}
  \includegraphics{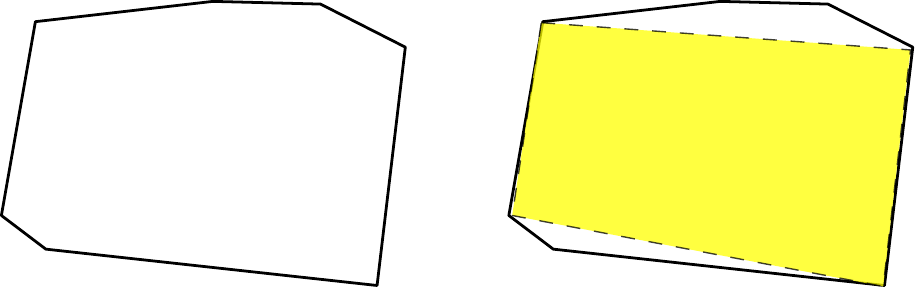}
  \centering
  \caption {(left) A convex polygon. (right) The largest-area inscribed quadrilateral.}
  \label {fig:example}
\end {figure}

 In this note, we revisit the following problem:  "Given a convex polygon $P$, what is the largest-area quadrilateral inscribed in $P$?" (see Figure~\ref{fig:example}). 
We show by  example that the presented algorithm~\cite{45} also fails to find the optimal solution for $k=4$. Our counter example is a polygon on 16 vertices in the range [0, 26500].
Our counter example, combined with the work in ~\cite{kluv}, \cite {48} and \cite{mmud} would suggest that the problem of finding the largest-area quadrilateral in linear time is still open. Although Boyce \etal~\cite{48} claimed one is presented by Shamos~\cite{shb}, but the cited manuscript cannot be found on-line.   In this note we show that this problem (and the problem for any constant value of $k$ ) can be solved in linear time.

Also after the initial posting of the manuscript~\cite{kluv} on arXiv, two new linear-time algorithms for solving the problem of finding the largest-area inscribed triangle have already been claimed  by \cite{kallus} and \cite{jin}.


\subsection {Definitions}

Let $P$ be a convex polygon with $n$ vertices.
Keikha \etal~\cite{kluv} define  a \emph {$P$-aligned} polygon $Q$ as  a convex polygon that is inscribed in $P$ with its vertices on vertices of $P$.
 Boyce \emph{et al.} \cite{48} define a \emph{rooted} polygon $Q$ with root $r \in P$ to be any $P$-aligned polygon that includes $r$.  Keikha \etal~\cite{45} define a vertex $v\in P$ of a polygon $Q$ stable, when moving $v$ reduces the area of $Q$ as long as we maintain the cyclic ordering of $Q$.
In the reminder, we denote by $\Lambda_{P,k}$  the largest (by area) $P$-aligned polygon with $k$ vertices. Also we denote $Q_{p,k}$ for $P$-aligned polygons with $k$ vertices. A polygon $Q$ is $k$-stable where it has  $k$ stable vertices.

\begin {figure}
  \includegraphics{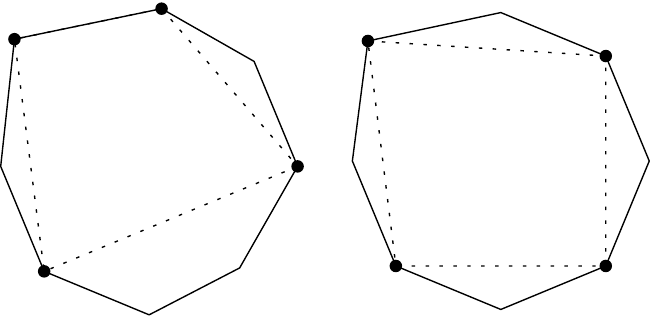}
   \centering
  \caption {Two 4-stable 4-gons inscribed in a convex polygon.}
  \label {fig:coin}
\end {figure}

 Note that all the vertices of $\Lambda_{P,k}$ are stable,  but a $k$-stable $Q_{P,k}$  does not necessarily coincides with the  $\Lambda_{P,k}$, as illustrated in Figure~\ref{fig:coin}.

Indeed the  idea of the presented method~\cite{45} was based on starting with a rooted $Q_{P,k}$ and moving the vertices of  $Q_{P,k}$ around the given polygon $P$ where keeping the cyclic ordering of $Q_{P,k}$ and increasing the area, and updating the area while finding a larger $k-1$-stable rooted polygon. 

This procedure will result in  keeping the sequence of the area of the potential solution only increasing. The authors~\cite{45} named this attribute as the \textit{unimodality of the area}, but we illustrated in Figure~\ref{fig:mmud} that keeping the unimodality will not result in finding the optimal solution necessarily.

\subsection {Dobkin and Snyder's  algorithm}
\begin {figure}
  \includegraphics{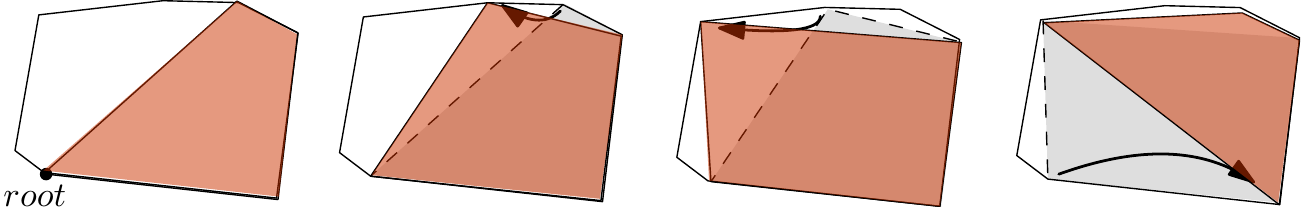}
   \centering
  \caption {The first three steps of Algorithm~\ref{alg:quadrangle}.}
  \label {fig:trialg}
\end {figure}

 We will now recall the \textit{quadrilateral  algorithm} \cite{45}, that is outlined in Algorithm~\ref {alg:quadrangle} and illustrated in Figure~\ref {fig:trialg}.

  Let $P=\{p_0,p_1,\ldots,p_{n-1}\}$. We assume that $P$ is given in a counter  clockwise orientation. Assume an arbitrary vertex of $P$ is is the root of the algorithm, assign this vertex and its three subsequent vertices  in the counter clockwise order on the boundary of $P$ to variables $a, b$, $c$ and $d$. 
  We then ``move $d$ forward'' along the  boundary of $P$ as long as this increases the area of $ abcd$.
  
If we can no longer advance $d$, we advance $c$ if this increases the area of $ abcd$, then try again to advance $d$.
If we can no longer advance $d$ and $c$, we advance $b$ if this increases the area of $ abcd$, then try again to advance $d$ and $c$.

 If we cannot advance either $d$, $c$ or $b$ any further, we advance $a$.
  We keep track of the largest-area quadrilateral found, and stop when $a$ returns to the starting position.
Since $a$ visits $n$ vertices and $d$, $c$ and $b$ each visit fewer than $2n$ vertices, the algorithm runs in $O(n)$ time (assuming we are given the cyclic ordering of the points on $P$).

Indeed, Algorithm~\ref{alg:quadrangle} is based on an observation that the largest inscribed quadrilateral treats as a unimodal function,which is not correct. 
 But, of course there is an  quadrilateral with some  vertices to be stable, that  Algorithm~\ref{alg:quadrangle} will find it in linear-time, but our counter example shows that the reported  quadrilateral does not necessarily equal to $\Lambda_{4,P} $. Furthermore, the reported quadrilateral is not even  4-stable.

\section{Counter-example to Algorithm~\ref {alg:quadrangle}}\label {sec:coex}

\begin {figure}
\includegraphics[scale=0.85]{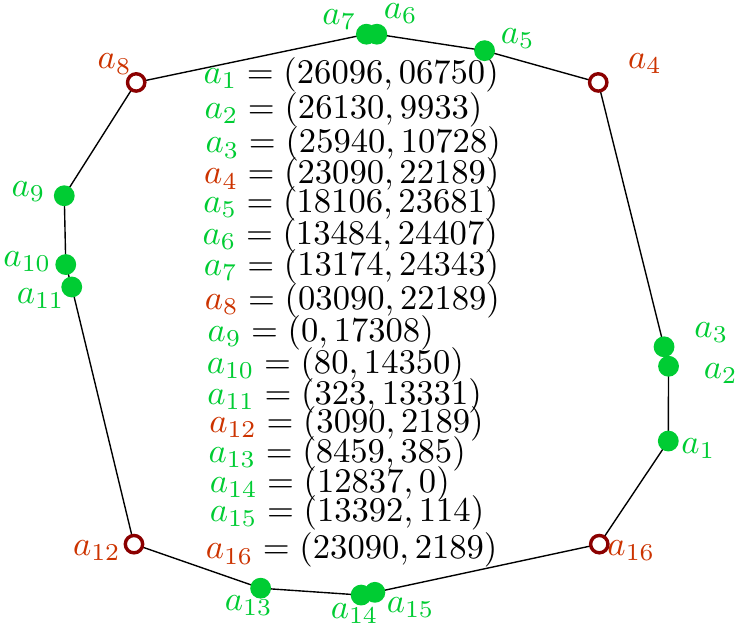}
\centering
\includegraphics[scale=0.85]{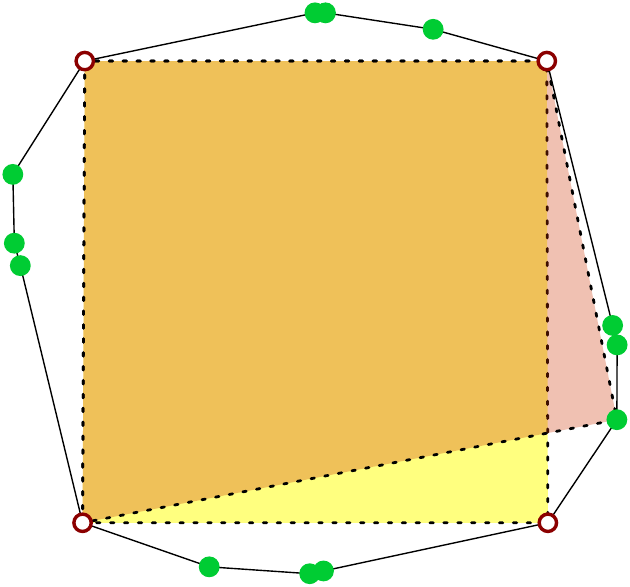}
\caption {(top) A polygon on $16$ vertices. (bottom)  The largest-area quadrilateral $a_4 a_8 a_{12} a_{16}$ (yellow), and the quadrilateral reported by Algorithm~\ref{alg:quadrangle}; $a_1 a_4 a_8 a_{12} $ (red).}
\label {fig:counter}
\end {figure}

 In  Figure~\ref{fig:counter} we provide a polygon $P$ on $16$ vertices such that $\Lambda_{4,P}$  and  the largest-area inscribed quadrilateral computed by Algorithm~\ref {alg:quadrangle} are not the same. 

\begin {figure}
\includegraphics{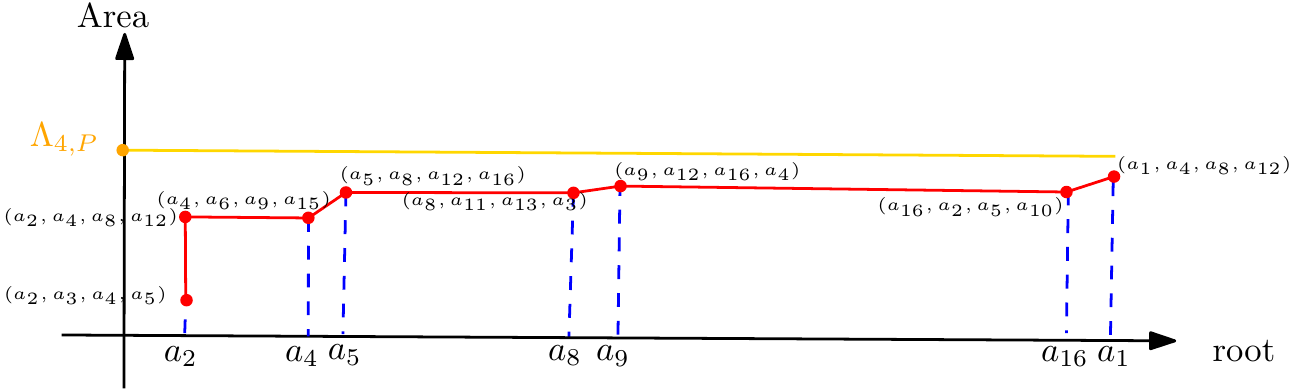}

\caption {Keeping the unimodality of the area of the potential solution during the algorithm will not result in the optimal solution necessarily.}
\label {fig:mmud}
\end {figure}
 We use the following points:
  $a_1=(26096,06750), a_2=(26130,9933), a_3=(25940,10728), a_4=(23090,22189), a_5=(18106,23681), a_6=(13484,24407),$ $ a_7=(13174,24343),  a_8=(3090,22189),  a_9=(0,17308),  a_{10}=(80,14350),  a_{11}=(323,13331),  a_{12}=(3090,2189),  a_{13}=(8459,385),  a_{14}=(12837 ,0),$ $ a_{15}=(13392,114) , a_{16}=(23090,2189)$. 
  The largest-area quadrilateral is $ a_4 a_8 a_{12} a_{16}$; however, Algorithm~\ref {alg:quadrangle} reports  $a_1  a_4 a_8 a_{12}$ as the largest-area quadrilateral, where starting the algorithm from any root. 
The results of running Algorithm~\ref{alg:quadrangle} where starting   on root $a_1$ are demonstrated on Figure~\ref{fig:mmud}.  
Thus,  the algorithm fails to find $\Lambda_{P,4}$ on any possible root.



\begin{algorithm} [H]
\SetAlgoLined

\caption{quadrilateral algorithm}


\label {alg:quadrangle}

     {\bf Input} {$P$: a convex polygon, $r$: a vertex of $P$}\\
     {\bf Output} { $m$: an quadrilateral}\\
     {\bf Legend} Operation {\texttt{\textit{next}} means the next vertex in counter clockwise order of $P$}\\
     a = r\\
     b = \texttt{\textit{next}}(a)\\
     c = \texttt{\textit{next}}(b)\\
    d = \texttt{\textit{next}}(c)\\
     m = \texttt{\textit{area}}($abcd$) \\
     
     \While{True}
{
       \While{\texttt{\textit{area}}($abcd$) $\leq$ \texttt{\textit{area}}(abc \texttt{\textit{next}}(d))}
       {
		d = \texttt{\textit{next}}(d)\;

 \While{\texttt{\textit{area}}($abcd$) $\leq$ \texttt{\textit{area}}(ab \texttt{\textit{next}}(c)d)}
       {
		 c = \texttt{\textit{next}}(c)\;

  }   
      \While{\texttt{\textit{area}}($abcd$) $\leq$ \texttt{\textit{area}}(a \texttt{\textit{next}}(b)cd)}
       {
		b = \texttt{\textit{next}}(b)\;
      }

}
      
         m = max(\texttt{\textit{area}}($abcd$),m)

        a = \texttt{\textit{next}}(a)\;

       \If{a=r}
    {
       \Return m\;
     }

 \If{b=a}
    {
       b = \texttt{\textit{next}}(b)\;
	 \If{c=b}
		 {c = \texttt{\textit{next}}(c)\;
		 \If{d=c}
 			{d = \texttt{\textit{next}}(d)\;}}
     }

 }   
\end{algorithm}

\bibliographystyle{elsarticle-num}


\bibliography{sample}

\end{document}